# Toroidal ferroelectricity in $PbTiO_3$ nanoparticles

M. G. Stachiotti, and M. Sepliarsky

*Instituto de Física Rosario, Universidad Nacional de Rosario, 27 de Febrero 210 Bis, (2000) Rosario, Argentina.*

We report from first-principles-based atomistic simulations that ferroelectricity can be sustained in $PbTiO_3$ nanoparticles of only a few lattice constants in size as a result of a toroidal ordering. We found that size-induced topological transformations lead to the stabilization of a ferroelectric bubble by the alignment of vortex cores along a closed path. These transformations, which are driven by the aspect ratio of the nanostructure, change the topology of the polarization field, producing a rich variety of polar configurations. For sufficiently flat nanostructures, a multi-bubble state bridges the gap between 0D nanodots and 2D ultra-thin films. The thermal properties of the ferroelectric bubbles indicate that this state is suitable for the development of nanometric devices.

Ferroic materials, ferromagnets, ferroelectrics and ferroelastics are similarly named for the reason that they all exhibit hysteretic responses to driving forces. For both ferroelectrics and ferromagnets, size effects are of keen interest for very practical reasons: their ferroic behavior makes them useful for storing information and the smaller they can be made the more information can be stored. Size effects on the ferroelectric phase transition in perovskite ultrafine particles strongly suggested the presence of a critical size, estimated on the order of hundreds of angstroms, below which the ferroelectric state becomes unstable. [1,2] Such behavior would render ferroelectrics useless for applications at sizes below a cutoff, thereby limiting their importance in future technologies. Low-dimensional ferroelectrics are also attractive from a fundamental point of view because they have revealed unique features intrinsic to the nanoscale size, like the formation of $180^o$ stripe domains in ultrathin films [3] and the prediction of vortex states in nanodots.[4] Ferroelectric materials generally form domain structures to reduce the depolarizing field energy. In very small ferroelectric systems, however, the formation of domain walls is not energetically favored, and a curling polarization configuration—that is, a vortex —has been proposed to occur. [4,5]

Vortices have been studied extensively in many physical systems, and specifically in magnetism, where they are associated with configurations of magnetic- ux closure and thus correspond to states of low magnetostatic energy. Vortex domain structures in magnetic materials were predicted by Kittel [6], who showed that formation of circular domains was likely in ferromagnetic nanodots due to the surface boundary conditions. Nevertheless, the experimental verification of vortex domain states in ferromagnets of small dimensions, as well as direct observations of out-of-plane magnetization inside the core region [7-8] and ferrotoroidic domains [9] has been a recent development in ferroic research. In the case of ferroelectrics, signatures of vortex states have been detected from switching data [10] and piezoresponse force microscopy [11,12], nevertheless an unambiguous observation of polarization vortices remains elusive.

The control of vortex configurations in nanoferroelectrics would open exciting opportunities for designing revolutionary devices. [13,14] For example, the posibility of vortex core alignment in nanodots has been discussed recently by Scott. [15] His arguments were based on the work of Gorbatsevich and Kopaev [16] who suggested the case where a ring-shaped torus is formed. Despite efforts, such a pattern has not been observed in nanoferroelectrics yet. [17-19] Here we demonstrate from first-principles-based atomistic simulations that size-induced topological transformations in nanodots lead to the generation of a ferroelectric bubble by the alignment of vortex cores along a closed path. These transformations, which are driven by the aspect ratio of the nanostructure, change the topology of the polarization field producing a rich variety of polar configurations. [20] When the vortex structure aligns by forming a ring-shaped torus, a ferroelectric bubble is developed as a strong resultant polarization concentrated through the center of the ring, directed perpendicular to the plane of the ring. The bubble is stabilized in flat rectangular $PbTiO_3$ nanodots with aspect ratio ~ 2.5, being the 8x8x3 unit cells nanoparticle the smallest one. Finally, we show that a multi-bubble pattern is generated when the nanoparticle is sufficiently flat. This state, characterized by the formation of stripe-shaped bubbles with alternating polarization, links 0D nanodots and 2D ultra-thin films.

The simulations were carried out using an atomistic model with parameters fitted to first-principles calculations. The validity of this approach was demonstrated by the accurate determination of the physical properties of a wide range of ferroelectrics. [21]

The particular model for $PbTiO_3$ reproduces correctly the cubic-tetragonal phase transition of the bulk [22], surface properties, such as relaxation patterns, relaxation energy and surface reconstruction [23], and interface effects of ultrathin films on $SrTiO_3$ in good agreement with experiments [24]. These precedents have motivated us to apply the same model to investigate nanoparticles through molecular dynamics simulations, which are carried out using the DL_POLY package [25]. We consider isolated stress-free $PbTiO_3$ nanodots under ideal open-circuit boundary conditions. The nanodots have rectangular shape, denoted by $N_x \times N_y \times N_z$ where $N_x$, $N_y$ and $N_z$ are the numbers of Ti atoms along the pseudocubic directions. All the faces consist of PbO planes. The local polarizations were calculated centered on Ti sites and our predicted patterns will be presented on a certain plane specified by its normal direction and its order index (e.g. x=5 is the 5th plane having a normal direction along the x axis). For all simulated nanoparticles, the total net polarization is found to be zero.

We analyze first the polarization pattern of cubic nanoparticles ($N_x = N_y = N_z = N$) for $N < 26$. Local polarizations are negligible for $N < 7$, while large ferroelectric off-center displacements exist for the larger nanodots. For instance, Figure 1(a) shows a representative $P_z$ polarization map for a 20×20×20 nanodot. Its interior is defined by two "domains" of equal and opposite polarization. However, the individual dipoles forming each domain are not aligned along the z direction, but they rotate from cell to cell forming a vortex-like pattern (Figure 1(b)). The ordering can be quantified by the toroidal moment $\mathbf{g} = 1/(2 N v) \; \Sigma \, \mathbf{r}_i \times \mathbf{p}_i$ [5] , where $\mathbf{p}_i$ is the dipole of cell i located at $\mathbf{r}_i$, N is the number of cells and $v$ is the cell volume. For all the investigated cases, the direction of $\mathbf{g}$ coincides with one of the <111> axes, indicating that the polarization curls around that direction. Figure 1(c) presents the temperature dependence of $g = (g_x^2$ $+g_y^2+g_z^2)^{1/2}$ for nanodots of different sizes. One can see that g is zero at high temperatures, while there is a temperature $T^*$ below which *g* grows when further reducing the temperature. $T^*$ is thus the highest temperature at which the dipoles adopt a vortex structure. For all the cases, T* is above room temperature indicating that the vortex state has a strong thermal stability.

The one-vortex configuration of the cubic nanodots transforms into a more complex state when the lateral size of the nanoparticle is increased. As example, we present results for thickness $N_z = 10$. The 10×10×10 nanodot displays a one-vortex state, as was described above. When $N_x$ and $N_y$ are slightly increased, a transformation into a four-vortex state takes place. The $P_z$ map for a 14×14×10 nanodot shows four domains, two pointing up and two pointing down (Figure 2(a)). The dipoles also rotate inside the nanodot, but now they are forming a four-vortex state. Two vortices have their toroidal moments lying along the <100> direction, opposite chirality, and cores located at the bottom half of the nanodot (Figure 2(b)). The remaining two have opposite toroidal moments along the <010> direction with their cores in the upper half (Figure 2(c)). The four vortices can be visualized together in Figure 2(d) where the pattern corresponding to a diagonal cut is presented.

When the lateral size overcomes some critical value, the four-vortex state transforms into a ring-shaped torus and a ferroelectric bubble is developed. Figure 3 shows the $P_z$ map (a) and a polarization pattern (b) for a 22×22×10 nanodot, where the bubble is perfectly formed. The polarization map displays a core domain state where the center of the nanodot is polarized up while the outer region polarizes down. The four vortices observed in the 14×14×10 nanodot are now aligned along a closed path, in the sense that a ring-shaped torus is created, generating a negative polarization in the regions nearby to the side-surfaces, and positive in the interior of the nanodot. A strong resultant polarization is then concentrated at the center of the ring. The profiles show that a bell-shaped ferroelectric bubble is formed, having a strong electric polarization aligned along the z axis, which is confined at the center of the torus. We note that ferroelectric bubble nanodomains were predicted in PZT ultrathin films, but as result of an incomplete screening. [26]

The size-induced topological transformation sequence described above is a general mechanism for the formation of ferroelectric bubbles in $PbTiO_3$ rectangular nanodots. We have repeated the simulations for nanodots with different values of $N_z$ and, in each case, bubbles appear when the nanodot aspect ratio $N_x$ / $N_z$ is greather than ~2.5 (see Figure 3(c)). We found that the 8×8×3 is the smallest nanoparticle containing a stable ferroelectric bubble. Note that despite the 3×3×3 nanodot does not present a vortex-like ground state, a bubble is generated due to the increment of the lateral size. Figure 3(d) indicates that the ferroelectric bubbles have a strong thermal stability. While the Curie temperature for the 8×8×3 nanodot is ~ 200K, the remaining bubbles are all stable at room temperature. Moreover, the 22×22×10 and 28×28×14 nanodots have Curie temperatures even higher than the bulk.

The evolution of the bubble state with the even greater increase in the lateral size of the nanodots leads to a particle-to-thin film crossover. Figure 4(a) shows the $P_z$ map and a polarization pattern for a 26×26×10 particle. Two bubbles with opposite polarization have been stabilized. Each bubble, however, does not display a circular shape; they are both elongated along the <100> direction. Vortices are detected in the regions nearby to the side-surfaces, while the region in between the bubbles (domain wall) presents an eddy structure of only two unit cell thick. If the lateral size of the nanodot becomes even larger, the number of bubbles increases. Figure 4(b) shows the patterns for a 40×40×10 nanodot where four stripe-shaped bubbles with alternating polarity form "$180^o$ stripe domains". This state resembles the domain configuration observed in $PbTiO_3$ ultrathin films. [3] For comparison, we show in Figure 4(c) the polarization pattern corresponding to an infinite slab of 10 unit-cell thick. It is interesting to note that the width of the stripe-shaped bubbles practically does not depend on the lateral size of the nanodots, coinciding with the width of the infinite strips generated in the film. These results lead to the reinterpretation of the stripe domain configuration observed in ultrathin films as an infinite succession of ferroelectric bubbles.

For completeness, we have calculated the polarization patterns of nanoparticles elongated along z ($N_x = N_y = 10 < N_z < 50$). When $N_z$ increases, the one-vortex state of the cubic nanodot transforms into another one-vortex state but with **g** parallel to the <001> direction. That is, the dipoles rotate around the z axis, in agreement with configurations found in ab-initio studies of $PbTiO_3$ nanowires.[27] Thus, the topological transformation path, from elongated nanorods to flat nanoplatelets, can be summarized as follows: one-vortex $\mathbf{g_{001}}$ → one-vortex $\mathbf{g_{111}}$ → four-vortex → torus (bubble) → multi-torus (multi-bubble). This topological sequence does not depend on the nature of the nanodot surfaces; the same result is obtained when all the faces of the nanodot consist of $TiO_2$ planes. Finally, we obtained that a ferroelectric bubble state can be stabilized also in a flat stress-free isolated nanodisk. In particular, we found a polar bubble in a disk of 25 unit-cell diameter and 10 unit-cell height. We note that core-polarization states were experimentally detected in cylindrical PZT nanodots by piezoresponse force microscopy. [11] Remarkably, the average aspect ratio of these nanodisks was 2.5, which agrees with our theoretical prediction.

The one-bubble state can be polarized equivalently parallel or anti-parallel to the z axis and it thus can be used as a ferroelectric bit. The properties showed in Figure 3 indicate that an array of nanobubbles is a promising device for ultrahigh-density recording, enabling a memory density well above the Tb/inch$^2$. The 22×22×10 nanoparticle, for example, develops a bubble of small dimensions, with large polarization (P=30μC/cm$^2$) and high Curie temperature ($T_c$ = 800K). The generation of a ferroelectric bubble by toroidal alignment may also have implications in the magnetoelectric coupling of multiferroic nanostructures. [14] Our results are thus interesting not only in terms of a fundamental knowledge of structural ordering in low-dimensional systems but also by their practical implications.

**Acknowledgements:** Supported by CONICET and ANPCyT de la República Argentina. M.G.S. thanks support from CIUNR.

FIGURE CAPTIONS

**Figure 1 (color online)** (a): $P_z$ polarization map (in μC/cm$^2$) for a 20×20×20 nanodot at T=50K. (b): Vortex-like structure of the local polarizations in a 20 × 20 × 20 nanoparticle. The figure corresponds to a cut along the x = 10th plane. The arrows give the direction of the projected displacement along the yz plane, and the arrow length indicates the projected magnitude. (c): Temperature dependence of the modulus of the toroidal moment for several cubic nanodots. The temperature has been scaled to match the experimental Curie temperature of the bulk.

**Figure 2 (color online)** (a): $P_z$ polarization map (in μC/cm$^2$) for a 14×14×10 nanodot at T=50K. (b-d): Local polarization patterns corresponding to three cuts along selected planes. In this nanoparticle the local polarizations rotate forming a four-vortex state.

**Figure 3 (color online)** (a): $P_z$ polarization map (in μC/cm$^2$) for a 22×22×10 nanodot at T=50K. (b): Local polarization patterns corresponding to a cut along a selected plane. The patterns indicate that a ferroelectric bubble is formed by vortex cores alignment along a closed path. The bubble displays a strong resultant polarization concentrated through the center of the ring, directed perpendicular to the plane of the ring. (c): Critical lateral size ($N_x$=$N_y$) where the bubble state is stabilized as a function of the nanodot thickness ($N_z$) . (d) Temperature dependence of the average polarization of the bubble for different bubble sizes. The temperature has been scaled to match the experimental Curie temperature of the bulk.

**Figure 4 (color online)**: $P_z$ polarization map (in μC/cm$^2$) and local polarization pattern for: (a) a 26×26×10 nanodot at T=50K where a bi-bubble state is stabilized, (b) a 40×40×10 nanodot at T=50K where four stripe-shaped bubbles with alternating polarity are formed. (c) Polarization pattern an infinite slab of 10 unit-cell thick at T=50K showing out-of-plane polarized nanostripes. The simulations for the slab were performed using periodic boundary condition along the x and y directions.

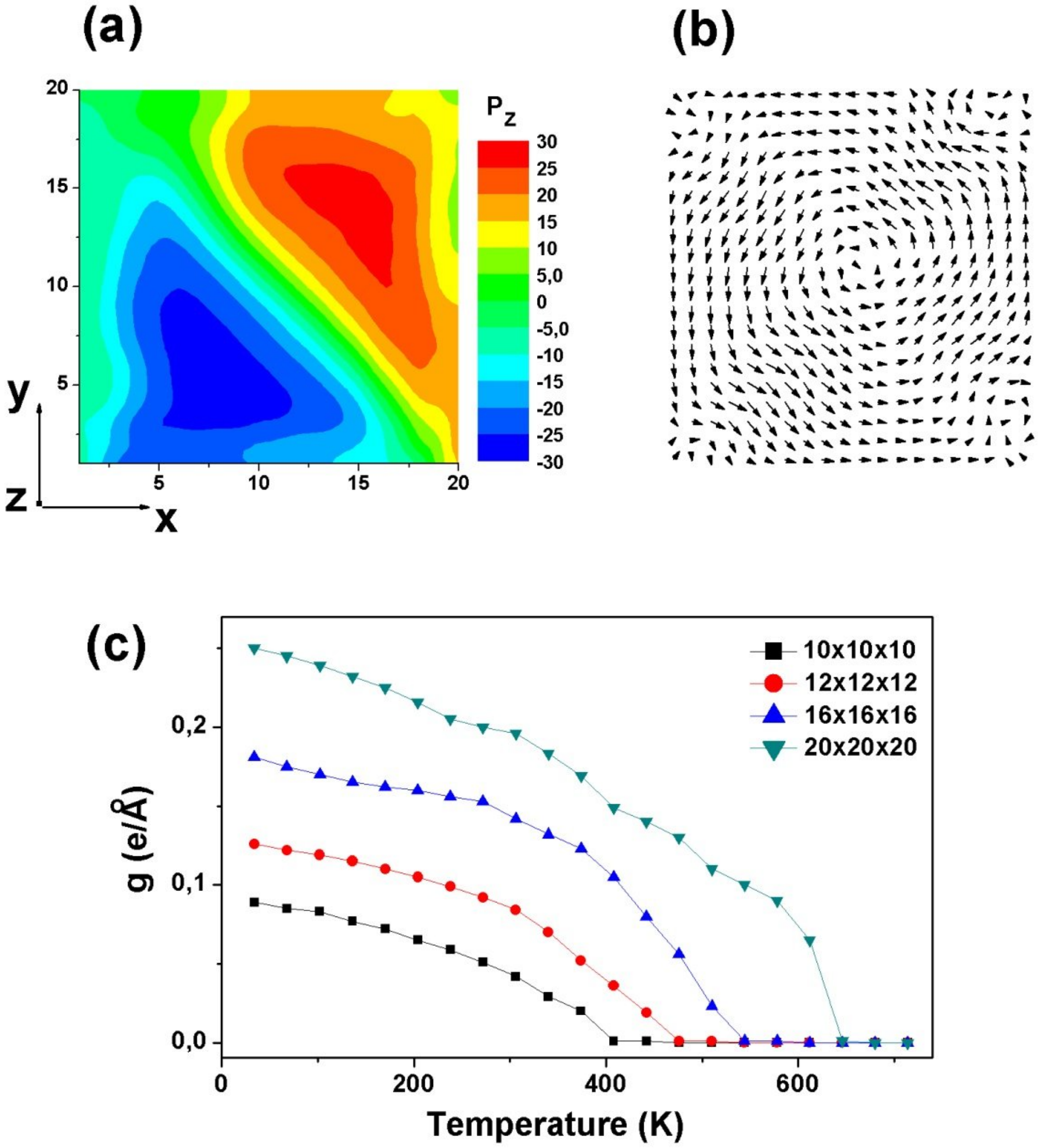

Figure 1

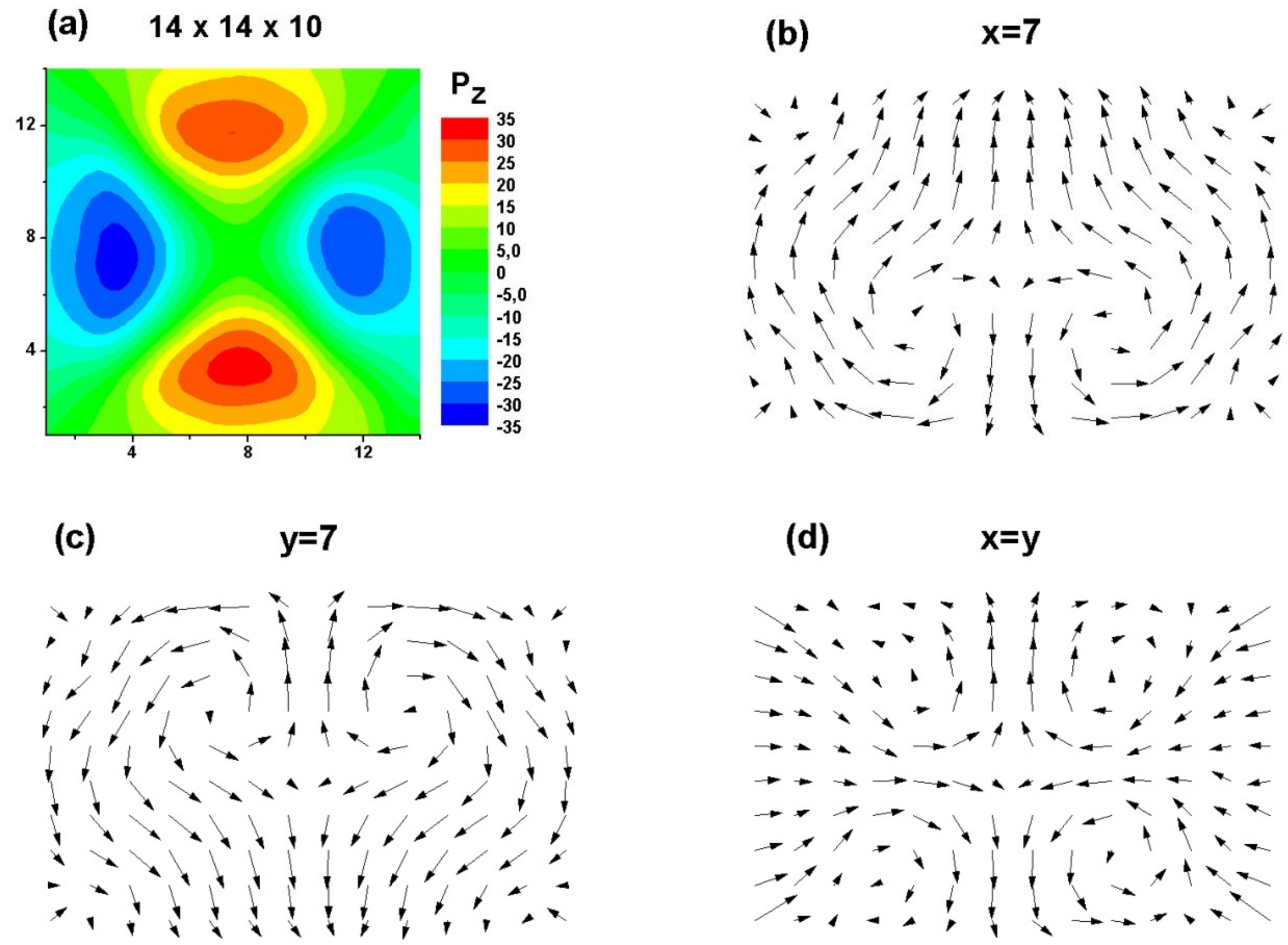

Figure 2

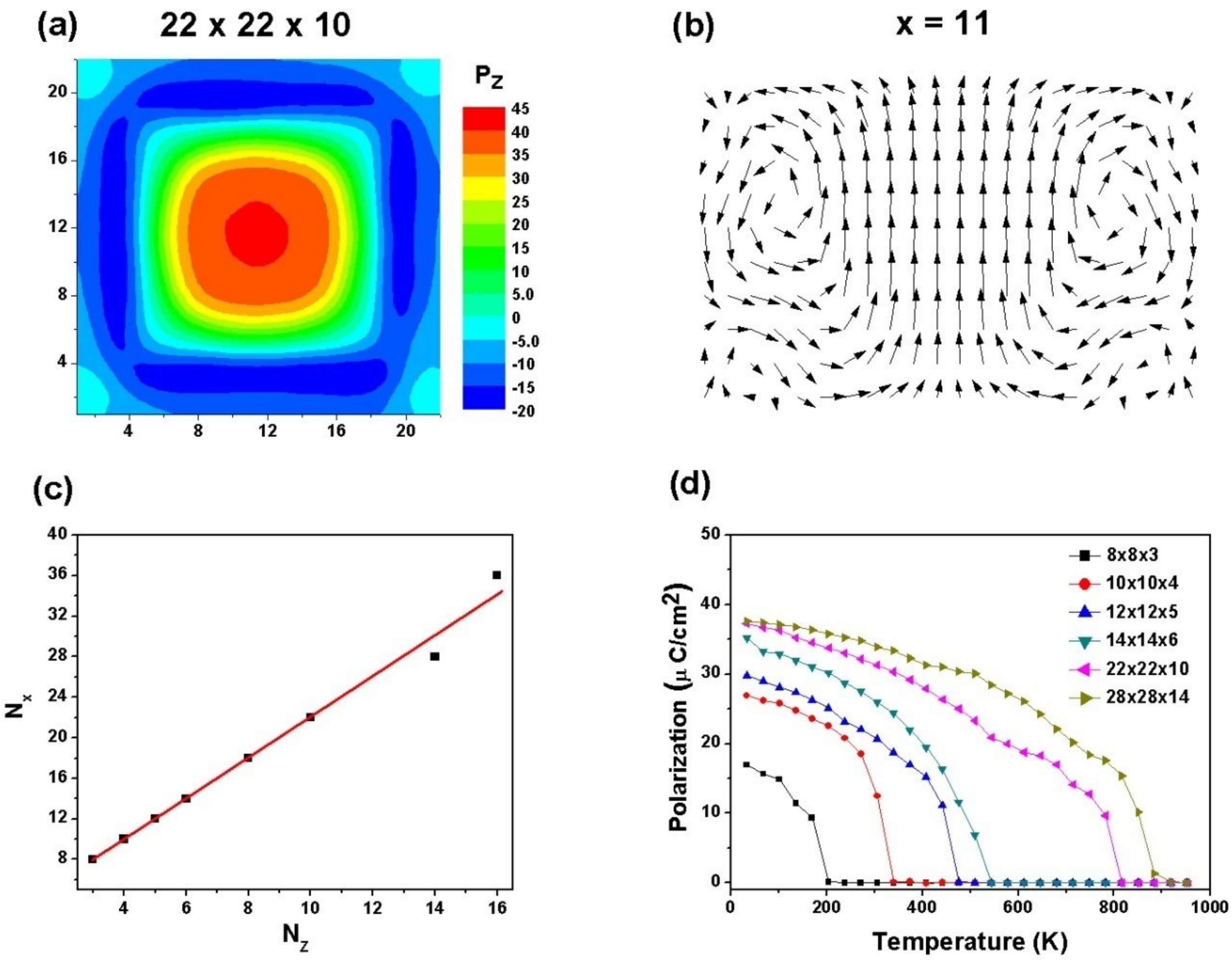

Figure 3

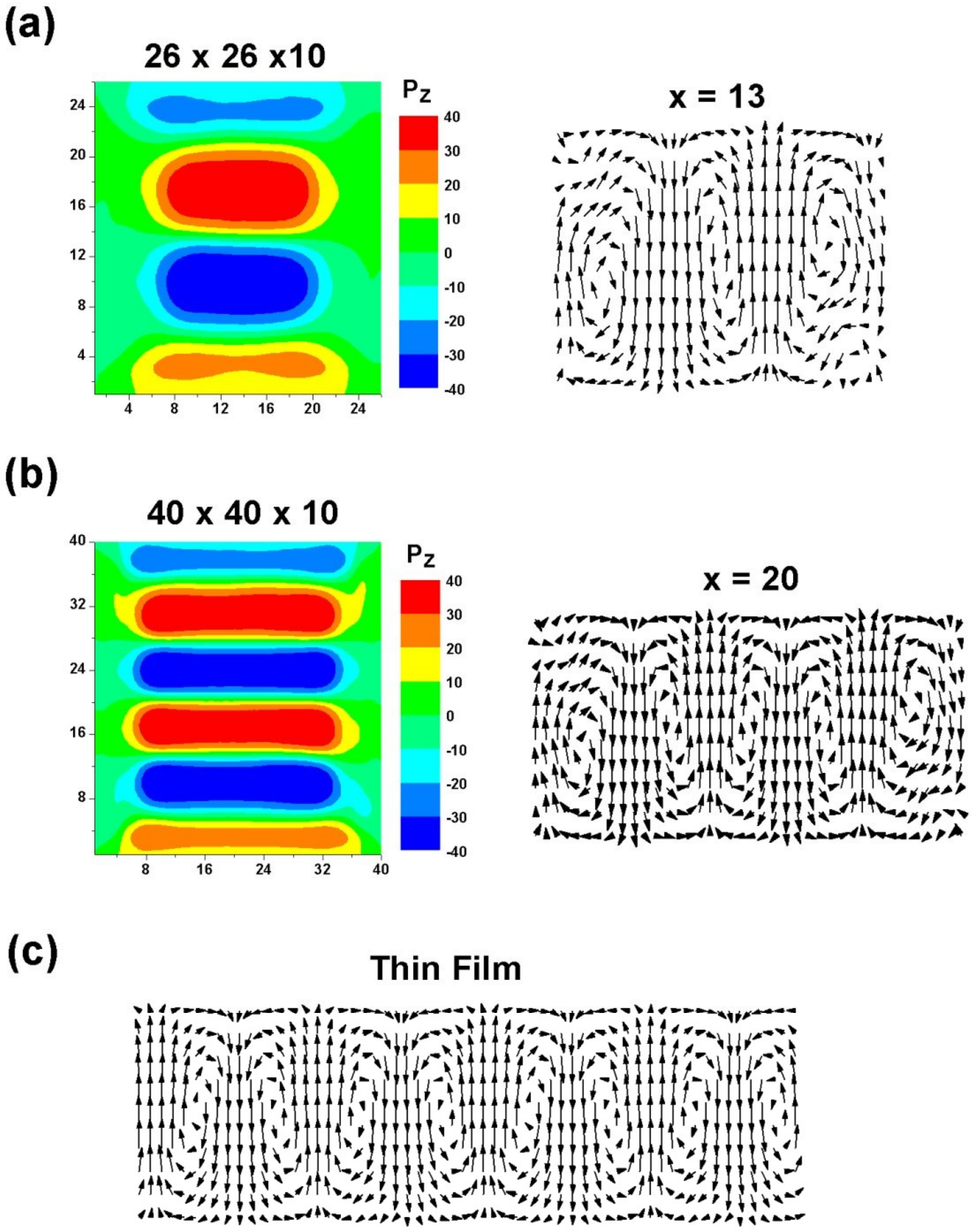

Figure 4